\documentclass[11pt]{article}
\usepackage{graphicx}
\usepackage{amsmath}
\usepackage{amssymb}
\usepackage{amsxtra}
\usepackage{bm}
\usepackage{mathrsfs}
\usepackage{cite}

\title{Ultraviolet divergences in supersymmetric theories regularized by higher derivatives}
\author{K.V. Stepanyantz\\Moscow State University, Faculty of Physics,\\ Department of Theoretical Physics,\\ 119991, Moscow, Russia,\\ stepan@m9com.ru}

\begin{document}
\maketitle

\begin{abstract}
Structure of quantum corrections in ${\cal N}=1$ supersymmetric gauge theories is investigated in the case of using the regularization by higher covariant derivatives. It is demonstrated that this regularization allows revealing some interesting features which lead to the exact relations between the renormalization group functions. In particular, the NSVZ equation, which relates the $\beta$-function to the anomalous dimension of the matter superfields, naturally appears in this case. We briefly review the all-loop derivation of this equation and the construction of a simple renormalization prescription under which it is valid.
\end{abstract}


\section{Introduction}\label{Section_Introduction}

The investigation of quantum corrections in supersymmetric theories is very important for both theory and phenomenological applications. As well known, ultraviolet (UV) divergences in these theories are restricted by some nonrenormalization theorems. The most known of them are the following:

1. ${\cal N}=1$ superpotential is not renormalized \cite{Grisaru:1979wc},

2. ${\cal N}=2$ theories are finite beyond the first loop \cite{Grisaru:1982zh, Howe:1983sr,Buchbinder:1997ib},

3. ${\cal N}=4$ supersymmetric Yang--Mills theory is all-loop finite \cite{Sohnius:1981sn,Grisaru:1982zh,Howe:1983sr,Mandelstam:1982cb,Brink:1982pd}.

\noindent
Due to the nonrenormalization theorems it is even possible to construct finite theories with ${\cal N}<4$ supersymmetry. For ${\cal N}=2$ supersymmetric theories this can be done by a special choice of a gauge group and  representations for the hypermultiplets \cite{Howe:1983wj}. For ${\cal N}=1$ supersymmetric theories \cite{Parkes:1984dh,Kazakov:1986bs,Ermushev:1986cu,Lucchesi:1987he,Lucchesi:1987ef} and theories with softly broken supersymmetry \cite{Jack:1997eh,Kazakov:1997nf,Kazakov:1995cy} it is also necessary to make a special tuning of a renormalization scheme.

However, the above list of the nonrenormalization theorems is not complete. For instance, it is reasonable to include into this list the exact Novikov, Shifman, Vainshtein, and Zakharov (NSVZ) $\beta$-function \cite{Novikov:1983uc,Jones:1983ip,Novikov:1985rd,Shifman:1986zi}. This is an equation which in general relates the $\beta$-function and the anomalous dimension of the matter superfields in ${\cal N}=1$ supersymmetric gauge theories,

\begin{equation}\label{Introduction_NSVZ}
\beta(\alpha,\lambda) = - \frac{\alpha^2 \Big(3 C_2 - T(R) + C(R)_i{}^j (\gamma_\phi)_j{}^i(\alpha,\lambda)/r\Big)}{2\pi(1-C_2 \alpha/2\pi)},
\end{equation}

\noindent
where $\alpha$ and $\lambda$ denote the gauge and Yukawa coupling constants, respectively, and we do not specify the definitions of the renormalization group functions (RGFs). The numerical factors in the equation (\ref{Introduction_NSVZ}) are defined as

\begin{eqnarray}
&& \mbox{tr}\,(T^A T^B) \equiv T(R)\,\delta^{AB};\qquad (T^A)_i{}^k (T^A)_k{}^j \equiv C(R)_i{}^j;\qquad\ \nonumber\\
&& f^{ACD} f^{BCD} \equiv C_2 \delta^{AB};\qquad\quad r \equiv \delta_{AA} = \mbox{dim}\, G.\qquad
\end{eqnarray}

However, there is an important problem, for which renormalization prescriptions the nonrenormalization theorems do hold. Really, the explicit calculations in ${\cal N}=1$ supersymmetric theories made in the $\overline{\mbox{DR}}$-scheme in \cite{Avdeev:1981ew,Jack:1996vg,Jack:1996cn,Harlander:2006xq} revealed that the NSVZ relation in this scheme is valid only in the one- and two-loop approximations, where this relation does not depend on a renormalization prescription. However, it turned out that with the help of a specially tuned finite renormalization of the gauge coupling constant one can restore the NSVZ equation, at least, in the three- and four-loop approximations \cite{Jack:1996vg,Jack:1996cn,Jack:1998uj,Harlander:2006xq,Mihaila:2013wma}. This (very nontrivial) fact implies that the NSVZ relation holds, but only in some special renormalization schemes usually called ``the NSVZ schemes''. Evidently, the $\overline{\mbox{DR}}$-scheme does not belong to them.

In this paper we briefly review the recent progress in constructing the all-loop prescription giving some NSVZ schemes based on the perturbative derivation of the NSVZ $\beta$-function made in \cite{Stepanyantz:2016gtk,Stepanyantz:2019ihw,Stepanyantz:2020uke}. The main ingredient needed for this is the regularization by higher covariant derivatives proposed by A.A.Slavnov \cite{Slavnov:1971aw,Slavnov:1972sq} rather long ago. Note that, by construction, it also includes the Pauli--Villars regularization for removing residual one-loop divergencies \cite{Slavnov:1977zf}. In the supersymmetric case this regularization can be self-consistently formulated in terms of ${\cal N}=1$ superfields \cite{Krivoshchekov:1978xg,West:1985jx} and, therefore, does not break supersymmetry.

\section{The NSVZ equation for ${\cal N}=1$ SQED}\label{Section_SQED}

\subsection{Regularization, quantization, and RGFs}\label{Subsection_SQED_RGFs}

As the simplest example illustrating how to derive the NSVZ equation and how to construct NSVZ schemes we consider the (massless) ${\cal N}=1$ supersymmetric electrodynamics (SQED) with $N_f$ flavors. Manifest supersymmetry at all steps of calculating quantum corrections is achieved by formulating the theory in terms of ${\cal N}=1$ superfields,

\begin{equation}\label{N=1_SQED_Action}
S = \frac{1}{4e_0^2}\mbox{Re}\int d^4x\,d^2\theta\,W^a W_a + \sum\limits_{\alpha=1}^{N_f} \frac{1}{4} \int d^4x\,d^4\theta\,\Big(\phi_\alpha^* e^{2V}\phi_\alpha + \widetilde\phi_\alpha^* e^{-2V} \widetilde\phi_\alpha\Big).
\end{equation}

\noindent
In the action (\ref{N=1_SQED_Action}) the (real) gauge superfield is denoted by $V$, while $\phi_\alpha$ and $\widetilde\phi_\alpha$ are chiral matter superfields with opposite $U(1)$ charges. In the Abelian case the strength of the superfield $V$ is defined as $W_a = \bar D^2 D_a V/4$.

Setting $C_2 = 0$, $C(R) = I$ (where $I$ is the $2N_f\times 2 N_f$ identity matrix), $T(R) = 2 N_f$, and $r=1$ in the equation (\ref{Introduction_NSVZ}) we see that in this case the NSVZ $\beta$-function takes the form \cite{Vainshtein:1986ja,Shifman:1985fi}\footnote{So far we again do not specify the definitions of RGFs.}

\begin{equation}\label{SQED_RGFs_Abelian_NSVZ}
\beta(\alpha) = \frac{\alpha^2N_f}{\pi}\Big(1-\gamma(\alpha)\Big)
\end{equation}

\noindent
and relates the $L$-loop $\beta$-function to the $(L-1)$-loop anomalous dimension of the matter superfields $\gamma(\alpha)$.

The considered theory will be regularized by higher covariant derivatives in two steps. First, we add a term with higher derivatives to the action (\ref{N=1_SQED_Action}) and regularize divergences beyond the one-loop approximation.  Then the resulting regularized action can be written as

\begin{eqnarray}
&& S_{\mbox{\scriptsize reg}} = \frac{1}{4 e_0^2} \mbox{Re}\,\int d^4x\,d^2\theta\,W^a R(\partial^2/\Lambda^2) W_a\nonumber\\
&& \qquad\qquad\qquad\quad + \sum\limits_{\alpha=1}^{N_f} \frac{1}{4} \int d^4x\,d^4\theta\,\Big(\phi_\alpha^* e^{2V}\phi_\alpha + \widetilde\phi_\alpha^*
e^{-2V} \widetilde\phi_\alpha\Big),\qquad
\end{eqnarray}

\noindent
where $R(\partial^2/\Lambda^2)$ is a regulator function, e.g., $R = 1 + \partial^{2n}/\Lambda^{2n}$. Next, for removing one-loop divergences and subdivergences that will be still present, the Pauli--Villars determinants are inserted into the generating functional,

\begin{equation}
Z[J,j,\widetilde j] = \int D\mu\, \Big(\det PV(V,M)\Big)^{N_f} \exp\Big\{i S_{\mbox{\scriptsize reg}} +iS_{\mbox{\scriptsize gf}} + S_{\mbox{\scriptsize sources}} \Big\}.
\end{equation}

\noindent
The masses of the corresponding (Pauli--Villars) superfields should satisfy the important condition $M = a \Lambda$ with $a\ne a(e_0)$.

Calculating RGFs it is important to distinguish between the ones defined in terms of the bare coupling constant $\alpha_0$,

\begin{equation}
\beta(\alpha_0) \equiv \frac{d \alpha_0(\alpha,\Lambda/\mu)}{d\ln\Lambda}\Big|_{\alpha=\mbox{\scriptsize const}};\qquad
\gamma(\alpha_0) \equiv -\frac{d \ln Z(\alpha,\Lambda/\mu)}{d\ln\Lambda}\Big|_{\alpha=\mbox{\scriptsize const}},
\end{equation}

\noindent
and the ones (standardly) defined in terms of the renormalized coupling constant $\alpha$ \cite{Kataev:2013eta},

\begin{equation}
\widetilde\beta(\alpha) \equiv \frac{d\alpha(\alpha_0,\Lambda/\mu)}{d\ln\mu}\Big|_{\alpha_0=\mbox{\scriptsize const}};\qquad
\widetilde\gamma(\alpha) \equiv \frac{d\ln Z(\alpha_0,\Lambda/\mu)}{d\ln\mu}\Big|_{\alpha_0=\mbox{\scriptsize const}}.
\end{equation}

\noindent
The former RGFs are independent of a renormalization prescription if a regularization is fixed, although they certainly depend on a regularization. The latter (standard) RGFs depend both on a regularization and on a renormalization prescription. However, both definitions of RGFs up to a formal replacement of the argument coincide in the so-called HD+MSL renormalization scheme \cite{Shakhmanov:2017wji,Stepanyantz:2017sqg},

\begin{equation}
\widetilde\beta(\alpha)\Big|_{\mbox{\tiny HD+MSL}} = \beta(\alpha_0\to \alpha); \qquad \widetilde\gamma(\alpha)\Big|_{\mbox{\tiny HD+MSL}} = \gamma(\alpha_0\to \alpha).
\end{equation}

\noindent
This implies that a theory is regularized by higher derivatives and only powers of $\ln\Lambda/\mu$ are included into the renormalization constants. (Because this way of removing divergences is very similar to minimal subtraction in the case of using the dimensional technique, it is called minimal subtractions of logarithms.)

\subsection{The three-loop $\beta$-function with the higher derivative regularization}

The first calculations of the lowest quantum corrections for ${\cal N}=1$ SQED regularized by higher derivatives revealed that with this regularization the integrals giving the $\beta$-function defined in terms of the bare coupling constant are integrals of total derivatives \cite{Soloshenko:2003nc} and even double total derivatives \cite{Smilga:2004zr} with respect to the momentum of the matter loop. The result for the three-loop $\beta$-function in this form can be found, e.g., in \cite{Stepanyantz:2011jy,Aleshin:2020gec}. Note that such integrals do not vanish due to singularities of the integrands,

\begin{equation}\label{Integral_Of_Double_Total}
\int \frac{d^4Q}{(2\pi)^4} \frac{\partial}{\partial Q^\mu} \frac{\partial}{\partial Q_\mu} \Big(\frac{f(Q^2)}{Q^2}\Big) = \frac{1}{4\pi^2} f(0) \ne 0,
\end{equation}

\noindent
where we assume that $f(Q^2)$ is a non-singular function rapidly decreasing at infinity. Using equations similar to (\ref{Integral_Of_Double_Total}) it is possible to reduce a number of loop integrations by 1. A detailed analyses revealed that this leads to the following simple graphical interpretation of the Abelian NSVZ equation \cite{Smilga:2004zr}. If one consider a vacuum supergraph, then a certain set of superdiagrams contributing to the $\beta$-function is produced by attaching two external gauge lines in all possible ways. From the other side, cuts of matter lines in the original vacuum supergraph give a set of two-point superdiagrams contributing to the anomalous dimension of the matter superfields, in which a number of loops is less by 1. The equation (\ref{SQED_RGFs_Abelian_NSVZ}) relates these two contributions. At the three-loop level a detailed check of this graphical interpretation has been done in \cite{Kazantsev:2014yna}. In particular, it turned out that for theories regularized by higher derivatives the NSVZ equation holds even at the level of loop integrals provided RGFs are defined in terms of the bare coupling constant. The explicit expressions for these RGFs found in \cite{Kazantsev:2020kfl} for an arbitrary function $R(x)$ are written as

\begin{eqnarray}
&&\hspace*{-4mm} \frac{\beta(\alpha_0)}{\alpha_0^2} = \frac{N_f}{\pi} + \frac{\alpha_0 N_f}{\pi^2} - \frac{\alpha_0^2 N_f}{\pi^3}\Big(N_f \ln a + N_f + \frac{N_f A}{2} + \frac{1}{2} \Big) +
O(\alpha_0^3);\nonumber\\
&&\hspace*{-4mm} \gamma(\alpha_0) = -\frac{\alpha_0}{\pi} +
\frac{\alpha_0^2}{\pi^2}\Big(N_f \ln a + N_f + \frac{N_f A}{2} + \frac{1}{2}\Big) + O(\alpha_0^3),
\end{eqnarray}

\noindent
where

\begin{equation}\label{SQED_Three-Loop_Regularization_Parameters}
A \equiv \int\limits_0^\infty dx\,\ln x\, \frac{d}{dx} \frac{1}{R(x)};\qquad a = \frac{M}{\Lambda}.
\end{equation}

\noindent
We see that they do not depend on the parameters fixing a subtraction scheme, but depend on the regularization parameters $A$ and $a$. However, for all their values the NSVZ equation is valid.

RGFs defined in terms of the renormalized coupling constant do not in general satisfy the NSVZ equation,

\begin{eqnarray}
&&\hspace*{-3mm} \frac{\widetilde\beta(\alpha)}{\alpha^2} = \frac{N_f}{\pi} + \frac{\alpha N_f}{\pi^2} - \frac{\alpha^2 N_f}{2\pi^3} - \frac{\alpha^2 N_f^2}{\pi^3}\Big( \ln a + 1 + \frac{A}{2} + b_2 - b_1 \Big) + O(\alpha^3)\nonumber\\
&&\hspace*{-3mm} \widetilde\gamma(\alpha) = -\frac{\alpha}{\pi} + \frac{\alpha^2}{2\pi^2} + \frac{\alpha^2 N_f}{\pi^2} \Big(\ln a + 1 + \frac{A}{2} - b_1 + g_1 \Big) + O(\alpha^3),
\end{eqnarray}

\noindent
and depend on the finite constants $b_i$ and $g_i$ fixing the renormalization prescription. These constants are defined by the equations

\begin{eqnarray}
&&\hspace*{-4mm} \frac{1}{\alpha_0} = \frac{1}{\alpha} - \frac{N_f}{\pi} \Big(\ln\frac{\Lambda}{\mu} + b_1\Big)
- \frac{\alpha N_f}{\pi^2} \Big(\ln\frac{\Lambda}{\mu} + b_2\Big) + O(\alpha^2).\nonumber\\
&&\hspace*{-4mm} Z = 1 + \frac{\alpha}{\pi}\Big(\ln\frac{\Lambda}{\mu}+g_1\Big) + O(\alpha^2).\qquad
\end{eqnarray}

\noindent
In the HD+MSL scheme they vanish

\begin{equation}
g_1 = b_1 = b_2 = 0,
\end{equation}

\noindent
and both definition of RGFs give the same functions up to the formal replacement of arguments. Evidently, in this case the NSVZ equation holds, although the scheme dependence is already essential.

The three-loop $\beta$-function and the two-loop anomalous dimension of the matter superfields in the $\overline{\mbox{DR}}$ and MOM schemes can be found in \cite{Jack:1996vg,Kataev:2013csa,Kataev:2014gxa}. In these schemes the NSVZ equation relating these RGFs does not hold.

\subsection{The all-loop results}

In \cite{Stepanyantz:2011jy} the all-loop expression for the $\beta$-function of ${\cal N}=1$ SQED (defined in terms of the bare coupling constant in the case of using the higher derivative regularization) was presented in the form of a functional integral over the gauge superfield $V$, which is an integral of double total derivatives in the momentum space. Certainly, this integral of double total derivatives is reduced to the sum of singular contributions with the help of equations similar to (\ref{Integral_Of_Double_Total}). The all-loop sum of the singularities which has also been calculated in \cite{Stepanyantz:2011jy} gives the exact NSVZ $\beta$-functions for RGFs defined in terms of the bare couplings,

\begin{equation}
\frac{\beta(\alpha_0)}{\alpha_0^2} = \frac{N_f}{\pi} \Big(1-\gamma(\alpha_0)\Big).
\end{equation}

\noindent
Thus, if ${\cal N}=1$ SQED is regularized by higher derivatives, then RGFs defined in terms of the bare coupling constant satisfy the NSVZ equation in all orders for an arbitrary renormalization prescription. (Note that both sides of this equation do not depend on the parameter $\xi_0$ in the gauge fixing term, so that this result is valid for an arbitrary $\xi$-gauge.) Analogous result has been obtained in \cite{Stepanyantz:2014ima} with the help of a different method.

Consequently, some NSVZ schemes for RGFs defined in terms of the renormalized coupling constant are given by the HD+MSL prescription. Note that they constitute a certain subclass in the continuous set of the NSVZ schemes described in \cite{Goriachuk:2018cac}. Really, minimal subtractions of logarithms can supplement various versions of the higher derivative regularization, which can differ in the form of the higher derivative regulator $R(x)$ and the constant $a$ defined by the equation (\ref{SQED_Three-Loop_Regularization_Parameters}). Note that the class of NSVZ schemes also includes the on-shell scheme, which is another all-loop NSVZ renormalization prescription \cite{Kataev:2019olb}.

\section{Non-Abelian supersymmetric gauge theories regularized by higher derivatives}\label{Section_SYM_HD}

Below we will describe the derivation of the NSVZ equation and NSVZ schemes for renormalizable non-Abelian ${\cal N}=1$ supersymmetric gauge theories. In the massless limit they are described by the classical action
\begin{eqnarray}
&&\hspace*{-5mm} S = \frac{1}{2 e_0^2}\,\mbox{Re}\,\mbox{tr}\int d^4x\, d^2\theta\,W^a W_a + \frac{1}{4} \int d^4x\, d^4\theta\,\phi^{*i}
(e^{2V})_i{}^j \phi_j\nonumber\\
&&\hspace*{-5mm}\qquad\qquad\qquad\qquad\qquad\qquad\quad + \Big(\frac{1}{6}\lambda_0^{ijk} \int d^4x\,d^2\theta\,  \phi_i \phi_j \phi_k + \mbox{c.c.}\Big),\qquad
\end{eqnarray}

\noindent
which is written in a manifestly supersymmetric form with the help of ${\cal N}=1$ superspace. Below we will assume that the gauge group $G$ is simple, and the chiral matter superfields $\phi_i$ transform under its representation $R$. The classical non-Abelian expression for the supersymmetric gauge superfield strength is $W_a = \bar D^2(e^{-2V} D_a e^{2V})/8$. Also the bare Yukawa couplings $\lambda_0^{ijk}$ should satisfy the condition

\begin{equation}
\lambda_0^{ijm} (T^A)_m{}^{k} + \lambda_0^{imk} (T^A)_m{}^{j} + \lambda_0^{mjk} (T^A)_m{}^{i} = 0,
\end{equation}

\noindent
which provides the gauge invariance of the cubic interaction term.

Quantizing the theory we use the background field method realized by the replacement $e^{2V} \to e^{2{\cal F}(V)} e^{2\bm{V}}$, where in the right hand side $\bm{V}$ and $V$ are the background and quantum gauge superfields, respectively. The function ${\cal F}(V)$ is needed for describing the nonlinear renormalization of the quantum gauge superfield \cite{Piguet:1981fb,Piguet:1981hh,Tyutin:1983rg} and includes an infinite set of parameters analogous to the parameter in the gauge fixing term. Explicit calculations \cite{Juer:1982fb,Juer:1982mp} demonstrated that this function is really different from $V$. Moreover, the renormalization group equations cannot be satisfied without taking the nonlinear terms into account \cite{Kazantsev:2018kjx}.

Following \cite{Aleshin:2016yvj,Kazantsev:2017fdc}, to introduce the regularization, we modify the action by adding terms with higher powers of the covariant derivatives

\begin{equation}
\nabla_a = D_a;\qquad \bar\nabla_{\dot a} = e^{2{\cal F}(V)} e^{2\bm{V}} \bar D_{\dot a} e^{-2\bm{V}} e^{-2{\cal F}(V)},
\end{equation}

\noindent
after which the regularized action takes the form

\begin{eqnarray}
&&\hspace*{-5mm} S_{\mbox{\scriptsize reg}} = \frac{1}{2 e_0^2}\,\mbox{Re}\, \mbox{tr} \int d^4x\, d^2\theta\, W^a \left(e^{-2\bm{V}} e^{-2{\cal F}(V)}\right)_{Adj} R\Big(-\frac{\bar\nabla^2 \nabla^2}{16\Lambda^2}\Big)_{Adj} \nonumber\\
&&\hspace*{-5mm} \times \Big(e^{2{\cal F}(V)}e^{2\bm{V}}\Big)_{Adj} W_a + \frac{1}{4} \int d^4x\,d^4\theta\, \phi^{*i} \Big[F\Big(-\frac{\bar\nabla^2 \nabla^2}{16\Lambda^2}\Big) e^{2{\cal F}(V)}e^{2\bm{V}}\Big]_i{}^j \phi_j
\nonumber\\
&&\hspace*{-5mm} + \Big(\frac{1}{6} \lambda_0^{ijk} \int d^4x\,d^2\theta\, \phi_i \phi_j \phi_k \Big) + \mbox{c.c.} \Big)
\end{eqnarray}

\noindent
and will contain the regulator functions $R(x)$ and $F(x)$. It is reasonable to choose the gauge fixing term which does not break the background gauge invariance and is analogous to the $\xi$-gauge fixing term in the usual Yang--Mills theory, although now it should contain one more regulator function $K(x)$. Certainly, the Faddeev-Popov and Nielsen--Kalosh ghosts must also be introduced according to the standard procedure.

The residual one-loop divergences are removed by inserting into the generating functional the Pauli--Villars determinants \cite{Slavnov:1977zf}. In the considered supersymmetric case one needs two such determinants \cite{Aleshin:2016yvj,Kazantsev:2017fdc},

\begin{eqnarray}
&& Z = \int D\mu\, \mbox{Det}(PV,M_\varphi)^{-1} \mbox{Det}(PV,M)^{T(R)/T(R_{\mbox{\tiny PV}})} \nonumber\\
&&\qquad\qquad\qquad\quad \times\exp\Big\{i\Big(S_{\mbox{\scriptsize reg}} + S_{\mbox{\scriptsize gf}} + S_{\mbox{\scriptsize FP}} + S_{\mbox{\scriptsize NK}} + S_{\mbox{\scriptsize sources}}\Big)\Big\},\qquad
\end{eqnarray}

\noindent
where the masses of the Pauli--Villars superfields $\varphi_{1,2,3}$ in the adjoint representation and $\Phi_i$ in a representation $R_{\mbox{\tiny PV}}$ are  $M_\varphi = a_\varphi \Lambda$ and $M = a\Lambda$, respectively, and the coefficients $a_\varphi$ and $a$ do not depend on couplings.

\section{The all-loop derivation of the NSVZ equation: the main steps}\label{Section_All_Loop_Derivation}

\subsection{The ultraviolet finiteness of the triple gauge-ghost vertices}\label{Subsection_VCC}

The first step needed for proving the non-Abelian NSVZ relation is the nonrenormalization theorem for the three-point vertices with two external lines of the Faddeev--Popov ghosts and one external line of the quantum gauge superfield \cite{Stepanyantz:2016gtk} (see also \cite{Korneev:2021zdz} for the generalization to the case of theories with multiple gauge couplings). According to this nonrenormalization theorem the triple gauge-ghost vertices are finite in all orders.\footnote{Similar statements in the Landau gauge were also proved for some theories formulated in terms of usual fields \cite{Capri:2014jqa,Dudal:2002pq}.} The all-loop proof of this statement in the general $\xi$-gauge is based on the superfield Feynman rules and the Slavnov--Taylor identities. The one- and two-loop checks of it have been done in \cite{Stepanyantz:2016gtk} and \cite{Kuzmichev:2021yjo,Kuzmichev:2021lqa}, respectively.

There are 4 vertices of the considered structure with the external lines corresponding to the superfields $\bar c\, V c$, $\bar c^+ V c$, $\bar c\, V c^+$, and $\bar c^+ V c^+$, where $\bar c$ and $c$ are the Faddeev--Popov antighost and ghost, respectively. The renormalization constants for all these vertices coincide and are equal to $Z_\alpha^{-1/2} Z_c Z_V$, where

\begin{equation}
\frac{1}{\alpha_0} = \frac{Z_\alpha}{\alpha};\qquad V = Z_V Z_\alpha^{-1/2} V_R;\qquad \bar c c = Z_c Z_\alpha^{-1} \bar c_R c_R,
\end{equation}

\noindent
so that the nonrenormalization theorem can be expressed by the equation

\begin{equation}\label{VCC_Nonrenormalization_Equation}
\frac{d}{d\ln\Lambda} (Z_\alpha^{-1/2} Z_c Z_V) = 0.
\end{equation}

\subsection{An equivalent form of the NSVZ equation}\label{Subsection_Equivalent_NSVZ}

The next step of the derivation is to rewrite the NSVZ relation in an equivalent form \cite{Stepanyantz:2016gtk}. For this purpose the non-Abelian NSVZ equation for RGFs defined in terms of the bare couplings is presented as

\begin{equation}\label{NSVZ_Rewritten}
\frac{\beta(\alpha_0,\lambda_0)}{\alpha_0^2} = - \frac{3 C_2 - T(R) + C(R)_i{}^j (\gamma_\phi)_j{}^i(\alpha_0,\lambda_0)/r}{2\pi}
+ \frac{C_2}{2\pi}\cdot \frac{\beta(\alpha_0,\lambda_0)}{\alpha_0}.
\end{equation}

\noindent
The $\beta$-function in the right hand side is expressed in terms of the anomalous dimensions $\gamma_c(\alpha_0,\lambda_0)$ and $\gamma_V(\alpha_0,\lambda_0)$ (of the ghosts and of the quantum gauge superfield, respectively) with the help of the equation (\ref{VCC_Nonrenormalization_Equation}),

\begin{eqnarray}
&& \beta(\alpha_0,\lambda_0) = \frac{d\alpha_0(\alpha,\lambda,\Lambda/\mu)}{d\ln\Lambda}\Big|_{\alpha,\lambda=\mbox{\scriptsize const}}
= -\alpha_0 \frac{d\ln Z_\alpha}{d\ln\Lambda}\Big|_{\alpha,\lambda=\mbox{\scriptsize const}}\nonumber\\
&&\qquad = -2\alpha_0 \frac{d\ln (Z_c Z_V)}{d\ln\Lambda}\Big|_{\alpha,\lambda=\mbox{\scriptsize const}}
= 2\alpha_0 \Big(\gamma_c(\alpha_0,\lambda_0) + \gamma_V(\alpha_0,\lambda_0)\Big).\qquad
\end{eqnarray}

\noindent
Substituting this expression into the the right hand side of (\ref{NSVZ_Rewritten}) we obtain the equation

\begin{eqnarray}\label{NSVZ_Equivalent_Form}
&& \frac{\beta(\alpha_0,\lambda_0)}{\alpha_0^2} = - \frac{1}{2\pi}\Big(3 C_2 - T(R) - 2C_2 \gamma_c(\alpha_0,\lambda_0)
\nonumber\\
&&\qquad\qquad\qquad\qquad - 2C_2 \gamma_V(\alpha_0,\lambda_0) + C(R)_i{}^j (\gamma_\phi)_j{}^i(\alpha_0,\lambda_0)/r\Big)\qquad
\end{eqnarray}

\noindent
relating the $\beta$-function to the anomalous dimensions of the quantum superfields. Note that the $\beta$-function in a certain loop is now expressed in terms of the anomalous dimensions only in the previous loop, while the original NSVZ equation relates the $\beta$-function only to the anomalous dimension of the matter superfields, but in all previous loops. That is why the NSVZ equation in the form (\ref{NSVZ_Equivalent_Form}) has a graphical interpretation analogous to the Abelian case. Namely, if one considers a supergraph without external lines, then attaching two legs of the backgrouns gauge superfield in all possible ways gives a contribution to $\beta(\alpha_0,\lambda_0)$, while various cuts of internal lines in the original vacuum supergraphs give contributions to the anomalous dimensions of the quantum superfields. The equation (\ref{NSVZ_Equivalent_Form}) relates all these contributions.

\subsection{The $\beta$-function and integrals of double total derivatives}\label{Subsection_Double_Integrals}

Next, it is necessary to prove that in supersymmetric theories regularized by higher covariant derivatives the $\beta$-function defined in terms of the bare couplings is given by integrals of double total derivatives with respect to the loop momenta.\footnote{This is not true for theories regularized by dimensional reduction due to a different structure of loop integrals \cite{Aleshin:2015qqc}.} This fact was observed in a large number of explicit calculations, see, e.g., \cite{Pimenov:2009hv,Stepanyantz:2011bz,Shakhmanov:2017soc,Kazantsev:2018nbl}. Its all-loop proof in the non-Abelian case has been done in \cite{Stepanyantz:2019ihw}. With the help of a rather complicated technique it was demonstrated that the $\beta$-function is determined by an expression which formally vanishes due to the Slavnov--Taylor identity corresponding to the background gauge invariance. Analyzing the Feynman rules it was demonstrated that that this expression is a sum of integrals of double total derivatives which are in fact nontrivial due to singularities of the integrands. Moreover, considering this formally vanishing expression as a starting point, one can construct a method for obtaining the $\beta$-function in ${\cal N}=1$ supersymmetric theories regularized by higher covariant derivatives in ${\cal N}=1$ superspace \cite{Stepanyantz:2019ihw,Stepanyantz:2019lyo,Kuzmichev:2019ywn}. This method requires to calculate only (specially modified) vacuum supergraphs and produces the result for a contribution to the $\beta$-function which comes from all superdiagrams obtained from them by attaching two external lines of the background gauge superfield. Note that this result is automatically obtained in the form of an integral of double total derivatives with respect to loop momenta.

The correctness of the method for calculating the $\beta$-function has been confirmed by a certain number of (very nontrivial) explicit calculations. For instance, using this method the two-loop $\beta$-function of ${\cal N}=1$ supersymmetric Yang--Mills theory with matter superfields in an arbitrary $\xi$-gauge has been reproduced in \cite{Stepanyantz:2019lyo}. In particular, it turned out that the NSVZ equations (\ref{Introduction_NSVZ}) and (\ref{NSVZ_Equivalent_Form}) are satisfied even at the level of loop integrals. However, in this approximation the NSVZ relations are scheme independent, so that it is desirable to consider the next order of the perturbation theory.

The explicit three-loop calculation is very complicated and (with the higher covariant derivative regularization) has not yet completely been done. However, a part of the three-loop $\beta$-function depending on the Yukawa couplings was found in \cite{Shakhmanov:2017soc,Kazantsev:2018nbl} with the help of the standard technique. Subsequently the same part of the $\beta$-function was obtained in \cite{Stepanyantz:2019ihw} with the help of the new technique which requires calculating only specially modified vacuum supergraphs. The results produced by both methods coincided. This means that the new method really works correctly.

\subsection{Summation of singular contributions}\label{Subsection_Songular_Sum}

The method for constructing integrals of double total derivatives described above can be used for deriving the exact NSVZ $\beta$-function in all orders of the perturbation theory. The main idea is that the integrals of double total derivatives can be taken with the help of equations similar to (\ref{Integral_Of_Double_Total}). Then they are reduced to the sums of singularities which correspond to various cuts of internal lines in vacuum supergraphs. These sums were calculated in all orders in \cite{Stepanyantz:2020uke} (see also \cite{Stepanyantz:2019lfm}). The result can be presented in the following form:

\begin{eqnarray}
&& \frac{\beta(\alpha_0,\lambda_0)}{\alpha_0^2} - \frac{\beta_{\mbox{\tiny 1-loop}}(\alpha_0)}{\alpha_0^2}
\\
&&= \frac{1}{\pi} C_2 \gamma_V(\alpha_0,\lambda_0) + \frac{1}{\pi} C_2 \gamma_c(\alpha_0,\lambda_0) - \frac{1}{2\pi r} C(R)_i{}^j (\gamma_\phi)_j{}^i(\alpha_0,\lambda_0).\qquad\nonumber
\end{eqnarray}
\begin{picture}(0,0.8)
\put(2,0.5){\vector(0,1){0.5}} \put(4.8,0.0){\vector(0,1){1.0}} \put(8.2,0.5){\vector(0,1){0.5}}
\put(0.7,0.1){gauge propagators}
\put(3,-0.4){Faddeev--Popov ghost propagators}
\put(7.2,0.1){matter propagators}
\end{picture}
\vspace*{8mm}

\noindent
We see that the sums of singularities corresponding to cuts of the gauge, ghost, and matter propagators produce the corresponding anomalous dimensions in the NSVZ equation written in the form (\ref{NSVZ_Equivalent_Form}). Substituting the one-loop expression for the $\beta$-function\footnote{With the higher covariant derivative regularization it has been calculated in \cite{Aleshin:2016yvj}.} we obtain the NSVZ equation for RGFs defined in terms of the bare couplings. Note that a renormalization prescription in this case is not fixed, but using of the higher covariant derivative regularization is highly essential.

Thus, we obtain the main result: the NSVZ relations (\ref{Introduction_NSVZ}) and (\ref{NSVZ_Equivalent_Form}) for RGFs defined in terms of the bare couplings are valid in all orders of the perturbation theory for RGFs defined in terms of the bare couplings if a theory is regularized by higher covariant derivatives. Note that these RGFs are scheme-independent for a fixed regularization, so that this result holds for any subtraction scheme supplementing the higher covariant derivative regularization.

\subsection{HD+MSL as an all-loop prescription giving some NSVZ schemes}\label{Subsection_Scheme_NSVZ}

In the HD+MSL scheme RGFs defined in terms of the renormalized couplings coincide with the ones defined in terms of the bare couplings. Therefore, from the above statement we obtain that the NSVZ equation (in both forms) for RGFs defined in terms of the renormalized couplings holds under the HD+MSL renormalization prescription in all orders of the perturbation theory. Note that this prescription gives a certain class of the NSVZ schemes,\footnote{A continuous set of NSVZ schemes was described in \cite{Goriachuk_Conference,Goriachuk:2020wyn,Korneev:2021zdz}.} because minimal subtractions of logarithms can supplement various versions of the higher covariant derivative regularization.

\section{Three-loop $\beta$-function for theories regularized by higher derivatives}\label{Section_Three_Loop_General}

Now we know a renormalization prescription under which the NSVZ equation holds. This allows to obtain a $\beta$-function in a certain loop by calculating the anomalous dimension of the matter superfields in the previous loops. For instance, the three-loop $\beta$-function for a general renormalizable ${\cal N}=1$ supersymmetric theory has been constructed in \cite{Kazantsev:2020kfl} with the help of the NSVZ equation and the expression for the two-loop anomalous dimension obtained with this regularization. Note that the NSVZ equation was used for RGFs defined in terms of the bare couplings, while the standard RGFs (also obtained in \cite{Kazantsev:2020kfl}) do not in general satisfy it. However, there is a certain particular case, for which the NSVZ equation in this order holds for an arbitrary renormalization prescription. This is theories finite in the one-loop approximation \cite{Parkes:1984dh} (see also \cite{Heinemeyer:2019vbc} for a recent review), which satisfy the conditions

\begin{equation}
T(R) = 3C_2; \qquad \lambda^*_{imn} \lambda^{jmn} = 4\pi \alpha C(R)_i{}^j.
\end{equation}

\noindent
Really, according to \cite{Kazantsev:2020kfl}, in this case the two-loop anomalous dimension of the matter superfields and the three-loop $\beta$-function defined in terms of the renormalized couplings have the form

\begin{eqnarray}\label{2-Loop_Gamma_Phi}
&&\hspace*{-5mm} (\widetilde{\gamma}_{\phi,2-loop})_i{}^j(\alpha,\lambda) = -\frac{3\alpha^2}{2\pi^2}C_2 C(R)_i{}^j\Big(\ln \frac{a_{\varphi}}{a} - b_{11} + b_{12}\Big)
- \frac{\alpha}{4\pi^2} \Big(\frac{1}{\pi} \lambda^*_{imn} \nonumber\\
&&\hspace*{-5mm}\qquad \times \lambda^{jml}  C(R)_l{}^n + 2\alpha \left[C(R)^2\right]_i{}^j\Big) \Big(A-B -2g_{12} +2g_{11}\Big); \\
\label{3-Loop_Beta}
&&\hspace*{-5mm}  \frac{\widetilde\beta_{3-loop}(\alpha,\lambda)}{\alpha^2} = \frac{3\alpha^2}{4\pi^3 r} C_2\, \mbox{tr}\left[C(R)^2\right] \Big(\ln \frac{a_\varphi}{a}  - b_{11} + b_{12}\Big)
+ \frac{\alpha}{8\pi^3 r} \Big(\frac{1}{\pi} C(R)_j{}^i \nonumber\\
&&\hspace*{-5mm}\qquad \times C(R)_l{}^n \lambda^*_{imn} \lambda^{jml} +2\alpha\,\mbox{tr}\left[C(R)^3\right]\Big) \Big(A-B -2g_{12}+2g_{11}\Big), \vphantom{\frac{1}{\pi^2}}
\end{eqnarray}

\noindent
where

\begin{equation}
A=\int\limits_0^\infty dx \ln x\, \frac{d}{dx}\frac{1}{R(x)};\quad\ B = \int\limits_0^\infty dx \ln x\, \frac{d}{dx}\frac{1}{F^2(x)}\quad\
a = \frac{M}{\Lambda};\quad\ a_\varphi = \frac{M_\varphi}{\Lambda},
\end{equation}

\noindent
and $b_i$ and $g_i$ are finite constants which fix a subtraction scheme in the lowest approximations. From the equations (\ref{2-Loop_Gamma_Phi}) and (\ref{3-Loop_Beta}) we see that for one-loop finite theories the NSVZ equation is satisfied in the lowest nontrivial approximation for an arbitrary renormalization prescription,

\begin{equation}
\frac{\beta_{3-loop}(\alpha,\lambda)}{\alpha^2} = - \frac{1}{2\pi r} C(R)_i{}^j (\gamma_{\phi,2-loop})_j{}^i(\alpha,\lambda).
\end{equation}

This result can be generalized. It is known that for ${\cal N}=1$ supersymmetric theories finite in the one-loop approximation one can tune a subtraction scheme so that the theory will be all-loop finite \cite{Kazakov:1986bs,Ermushev:1986cu,Lucchesi:1987he,Lucchesi:1987ef}. If a subtraction scheme is tuned in such a way that the $\beta$-function  vanishes in the first $L$ loops and the anomalous dimension of the matter superfields vanishes in the first $(L-1)$ loops, then \cite{Stepanyantz:2021dus} for an arbitrary renormalization prescription the $(L+1)$-loop gauge $\beta$-function satisfies the relation

\begin{equation}
\frac{\beta_{L+1}(\alpha,\lambda)}{\alpha^2} = - \frac{1}{2\pi r} C(R)_i{}^j (\gamma_{\phi,L})_j{}^i(\alpha,\lambda).
\end{equation}

\noindent
This implies that if a theory is finite in a certain approximation, then its $\beta$-function vanishes in the next order in exact agreement with the earlier known result of \cite{Parkes:1985hj,Grisaru:1985tc}.

\section{Conclusion}

Using the regularization by higher covariant derivatives, it is possible to reveal a number of interesting quantum features in supersymmetric theories. In particular, it is possible to construct an all-loop perturbative proof of the NSVZ equation and formulate simple renormalization prescriptions under which it is valid. Namely, it is valid for RGFs defined in terms of the bare couplings for any subtraction scheme supplementing this regularization. This fact has a simple graphical interpretation and follows from the factorization of loop integrals giving the $\beta$-function into integrals of double total derivatives, which takes place with the higher covariant derivative regularization. (Also the proof involves the nonrenormalization of triple gauge-ghost vertices, which is another interesting feature of quantum corrections in supersymmetric theories.) The usual RGFs (defined in terms of the renormalized couplings) satisfy the NSVZ equation in the HD+MSL scheme, when in theories regularized by higher covariant derivatives minimal subtractions of logarithms are used for removing divergences. This statement sheds light on the question of how one should calculate quantum corrections to obtain some exact result in various supersymmetric theories.

\section*{Acknowledgements}

This work has been supported by Foundation for Advancement of Theoretical Physics and Mathematics ``BASIS'', grant  No. 19-1-1-45-1.

\end{document}